\documentclass[twocolumn,prb]{revtex4}
\usepackage{amsfonts}
\usepackage[T1]{fontenc}
\usepackage{amsmath,amsbsy,amssymb,graphicx}
\usepackage{times}

\let\mathbf=\boldsymbol

\begin{document}

\title{Braiding of Majorana corner states in electric circuits and its non-Hermitian generalization}
\author{Motohiko Ezawa}
\affiliation{Department of Applied Physics, University of Tokyo, Hongo 7-3-1, 113-8656,
Japan}

\begin{abstract}
We propose to realize Majorana edge and corner states in electric
circuits. First, we simulate the Kitaev model by an LC electric circuit and
the $p_{x}+ip_{y}$ model by an LC circuit together with operational
amplifiers. Zero-energy edge states emerge in the topological phase, which are
detectable by measuring impedance. Next, we simulate the
Bernevig-Hughes-Zhang model by including an effective magnetic field without
breaking the particle-hole symmetry, where zero-energy corner states emerge in
the topological phase. It is demonstrated that they are Ising anyons subject
to the braiding. Namely we derive $\sigma ^{2}=-1$ for them, where $\sigma $
denotes the single-exchange operation. 
They may well be called Majorana states.
We also study non-Hermitian
generalizations of these models by requiring the particle-hole symmetry. It
is shown that the braiding holds in certain reciprocal non-Hermitian
generalizations.
\end{abstract}

\maketitle

\textit{Introduction:\ } A Majorana state will be a key for future
topological quantum computations\cite{TQC} owing to the braiding. Majorana
states are realized in topological superconductors\cite{Qi,Sato} and Kitaev
spin liquids\cite{Kitaev,Matsuda}. In these systems, the particle-hole
symmetry (PHS) plays an essential role since the zero-energy states becomes
Majorana states\cite{Alicea,Been,Elli}. Recently, Majorana corner states are
proposed in various systems\cite{ZYan,QWang,Vol,YWang,TLiu,Zhu,Pahomi},
where the braiding has been shown for some of them\cite{Zhu,Pahomi}.
Furthermore, Majorana states in non-Hermitian systems are studied in various contexts\cite%
{Malzard,SanJose,YuceM,Li2,Li,Coba,Zyuzin,Avila,Katsura,kawabata}. It is an
interesting problem to seek other systems realizing Majorana states.
Especially, it is fascinating if Majorana states are simulated by electric
circuits.

Various topological phases are realized in electric circuits, such as the
SSH model\cite{ComPhys}, graphene\cite{ComPhys,Hel}, Weyl semimetal\cite%
{ComPhys,Lu}, nodal-line semimetal\cite{Research,Luo}, higher-order
topological phases\cite{TECNature,Garcia,EzawaTEC}, Chern insulators\cite%
{Hofmann} and non-Hermitian topological phases\cite{EzawaLCR,EzawaSkin}. A
topological phase transition is induced by tuning variable capacitors and
inductors. The impedance signals the edge and corner states\cite%
{TECNature,ComPhys,Garcia,Hel,Hofmann,EzawaTEC,EzawaLCR,EzawaSkin}.

In this paper, we propose to realize Majorana edge states and corner states
in electric circuits. The essence is that parameters are tunable to make the
system respect the PHS. First, we simulate the Kitaev model either by a pure
LC circuit or by an LC circuit together with operational amplifiers. The $%
p_{x}+ip_{y}$ model is also constructed by aligning these two circuits along
the orthogonal directions. Zero-energy edge states corresponding to Majorana
states are well observed by measuring the impedance. Next, we \
simulate the Bernevig-Hughes-Zhang (BHZ) model together with an effective
Zeeman field by an electric circuit. The model corresponds to a second-order
topological superconductor with the emergence of a pair of Majorana corner
states. We demonstrate explicitly that the braiding holds between a pair of corner states by
calculating the Berry phase in an electric circuit.

The introduction of resistance makes the electric circuit non-Hermitian due
to the Joule loss\cite{EzawaLCR}. There are two types of non-Hermitian
models, i.e., reciprocal models and nonreciprocal models, where
nonreciprocity indicates that the forward and backward hopping amplitudes
are different between two nodes. As far as the PHS is respected, it is shown
that the braiding holds for the two corner states in a reciprocal
non-Hermitian extension of the BHZ model.
However, the braiding {becomes meaningless in the nonreciprocal extension.

\begin{figure}[t]
\centerline{\includegraphics[width=0.48\textwidth]{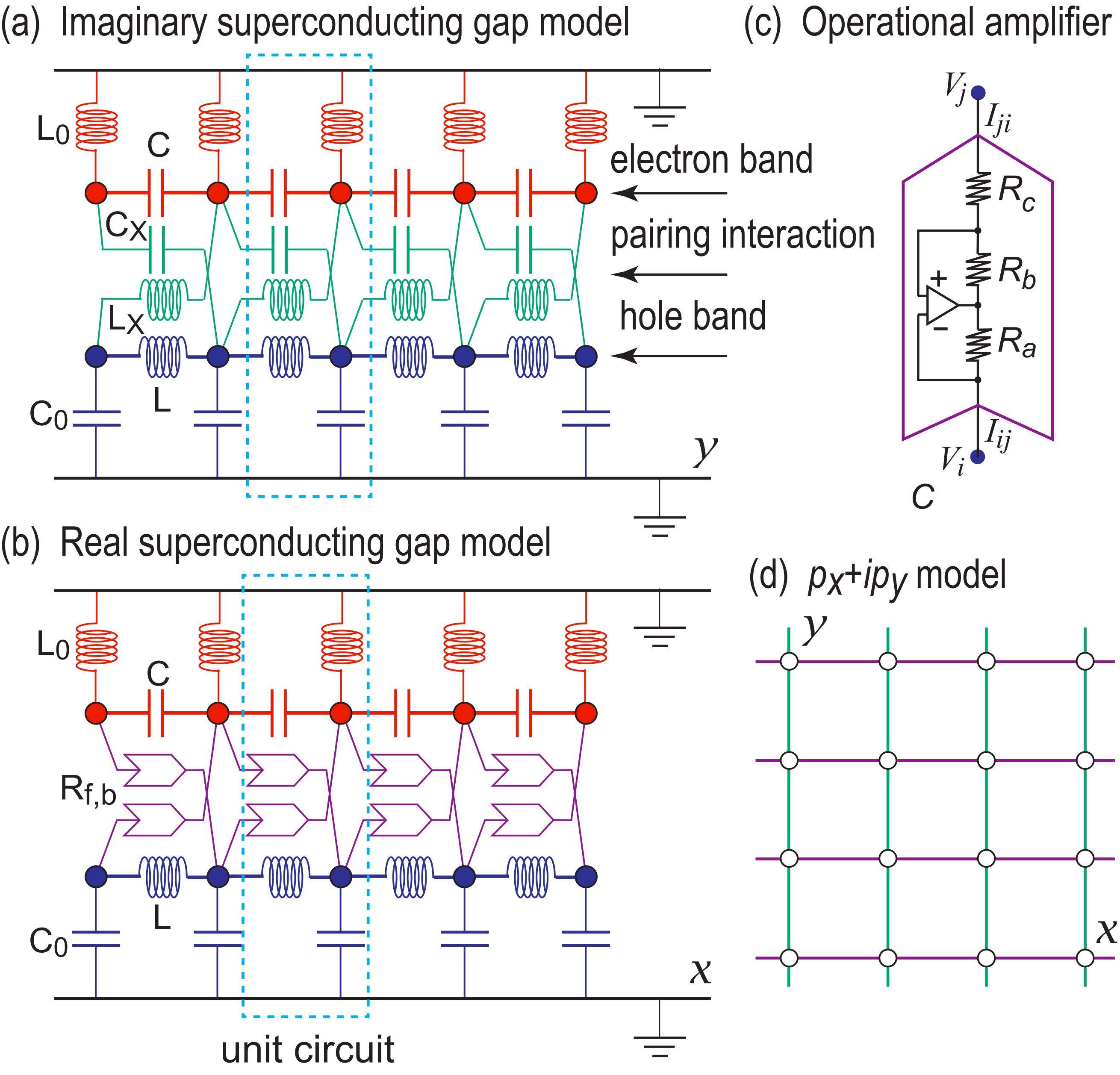}}
\caption{ Minimal electronic circuits realizing the Kitaev model.
Each circuit consists of two main wires colored in red
and blue, simulating electron and hole bands, respectively. The unit cell is
shown by a dotted cyan box, and contains two nodes colored in red and blue.
(a) Imaginary superconducting gap model is realized by a pure LC circuit.
(b) Real superconducting gap model is realized by an additional use of
operational amplifiers. (c) Structure of an operational amplifier. (d) The $%
p_{x}+ip_{y}$ model is constructed so that the 1D circuits (a) and (b) are
aligned in the $x$ and $y$ directions, respectively. Green and purple links
represent pairing interactions of the types (a) and (b), respectively. }
\label{FigCircuit}
\end{figure}

\textit{Electric-circuit realization of the Kitaev model:\ }The Kitaev $p$%
-wave topological superconductor model is the fundamental model hosting
Majorana zero-energy edge states in one dimensional (1D) space. The model is
represented\ in the two forms, i.e., by the Hamiltonian $H^{y}$ with the
imaginary superconducting pairing, and $H^{x}$ with the real superconducting
pairing. It consists of the hopping term $H_{t}$ and the superconducting
interaction term $H_{\text{SC}}^{i}$, where 
\begin{equation}
H^{i}=H_{t}\sigma _{z}+H_{\text{SC}}^{i},  \label{KitaHami}
\end{equation}%
with%
\begin{eqnarray}
H_{t} &=&-t\cos k-\mu ,  \label{KitaHop} \\
H_{\text{SC}}^{y} &=&\Delta _{y}\sigma _{y}\sin k,\qquad H_{\text{SC}%
}^{x}=\Delta _{x}\sigma _{x}\sin k.  \label{KitaSO}
\end{eqnarray}%
Here, $t$, $\mu $\ and $\Delta _{i}$\ represent the hopping amplitude, the
chemical potential and the superconducting gap parameter. It is a two-band
model, and the Hamiltonian is a $2\times 2$ matrix. It is well known that
the system is topological for $\left\vert \mu \right\vert <\left\vert
2t\right\vert $ and trivial for $\left\vert \mu \right\vert >\left\vert
2t\right\vert $ irrespective of $\Delta _{i}$ provided $\Delta _{i}\neq 0$.

We simulate the Kitaev model by electric circuits. There are two types of
circuits corresponding to the two Hamiltonians $H^{y}$ and $H^{x}$, as
illustrated in Fig.\ref{FigCircuit}(a) and (b). 

Let us explain how to construct them.
We use two main wires to represent a two-band model: One wire consists of capacitors $C$
in series, implementing the electron band, while the other wire consists of
inductors $L$ in series, implementing the hole band. The hopping parameters
are opposite between the electron and hole bands, which are represented by
capacitors and inductors. Indeed, they contribute the terms proportional to $%
i\omega C$ and $1/(i\omega L)$ to the circuit Laplacian $J_{ab}(\omega )$ in
(\ref{CircuLap}), respectively, where $\omega $ is the frequency of the $AC$ current. 

In the wire with capacitors (inductors), each node $a$ is connected to the
ground via an inductor $L_{0}$ ($C_{0}$), as in Fig.\ref{FigCircuit}(a)--(b). 
This setting is made to make the system topological.

\begin{figure}[t]
\centerline{\includegraphics[width=0.48\textwidth]{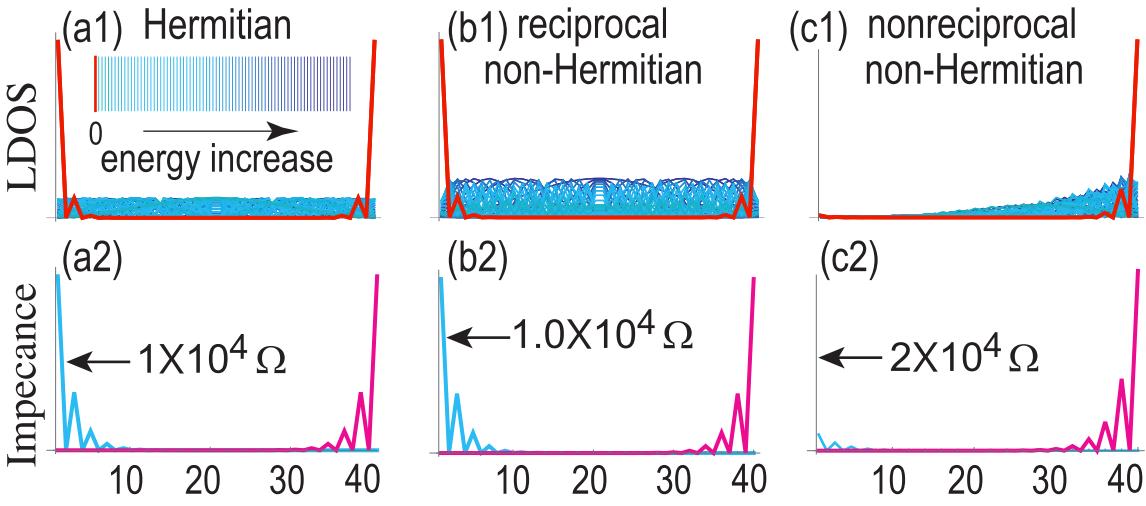}}
\caption{ (a1)--(c1) LDOS of a finite chain of the Hermitian and
non-Hermitian Kitaev models with length $N=40$ in the topological phase,
where the red peaks at the edges represent zero-energy edge states. Skin states
are observed in the nonreciprocal system (c1). The energy of the state is
represented by a color subject to the color palette in the inset of (a1).
(a2)--(c2) Two-point impedance of the corresponding LC circuit with
operational amplifiers, where the peaks at the edges are due to zero-energy
edge states. The horizontal axis is the site index. Magenta (cyan) curves
indicate the impedance when one node is fixed at the left (right)
edge. The parameters are as follows: $t=1$, $\Delta=0.5$ and $\protect\mu =0$
for the Kitaev model (a); $t=1+0.2i$, $\Delta=0.5$ and $\protect\mu =0$ for
the reciprocal non-Hermitian Kitaev model (b); $t=1$, $\Delta^f=0.5$, $%
\Delta ^{b}=0.3$ and $\protect\mu =0$ for the nonreciprocal non-Hermitian
Kitaev model (c). }
\label{FigKitaLDOS}
\end{figure}

We then introduce pairing interactions between them. In order to construct
the model $H^{y}$, we cross bridge two wires by capacitors $C_{X}$ and
inductors $L_{X}$ as shown in Fig.\ref{FigCircuit}(a). On the other hand, in
order to construct the model $H^{x}$, we cross bridge two wires by
operational amplifiers, which act as negative impedance converters with
current inversion\cite{Hofmann}. In the operational amplifier, the
resistance depends on the current flowing direction; $R_{f}$ for the forward
flow and $-R_{b}$ for the backward flow with the convention that $R_{b}>0$.
We set $R_{X}=R_{f}=R_{b}$ to make the system reciprocal. A generalization
to the nonreciprocal theory with $R_{f}\neq R_{b}$\ is straightforward.

The unit cell indicated by a dotted cyan box contains two sites in Fig.\ref{FigCircuit}.
Accordingly, we set $I_{a}=(I_{a}^{L},I_{a}^{C})$ and $V_{a}=(V_{a}^{L},V_{a}^{C})$, 
where $I_{a}^{L(C)}$ is the current between node $a^{L(C)}$ and the ground via
the inductance $L$ (conductance $C$), and $V_{a}^{L(C)}$ is the voltage at node $a^{L(C)}$.  

When we apply an AC voltage $V\left( t\right) =V\left( 0\right) e^{i\omega
t} $, the Kirchhoff current law leads to the following formula\cite{ComPhys,TECNature}, 
\begin{equation}
I_{a}\left( \omega \right) =\sum_{b}J_{ab}\left( \omega \right) V_{b}\left(
\omega \right) ,  \label{CircuLap}
\end{equation}%
where the sum is taken over all adjacent nodes $b$, and $J_{ab}\left( \omega
\right) $ is called the circuit Laplacian.

(i) 
In the case of the circuit in Fig.\ref{FigCircuit}(a) we explicitly obtain
\begin{equation}
J=\left( 
\begin{array}{cc}
f_{1} & g_{1} \\ 
g_{2} & f_{2}%
\end{array}%
\right) ,  \label{MatrixJ}
\end{equation}%
with 
\begin{align}
f_{1}& =-2C\cos k+2C-(\omega ^{2}L_{0})^{-1},  \notag \\
f_{2}& =2(\omega ^{2}L)^{-1}\cos k-2(\omega ^{2}L)^{-1}+C_{0},  \notag \\
g_{1}& =-C_{X}e^{ik}+(\omega ^{2}L_{X})^{-1}e^{-ik},  \notag \\
g_{2}& =(\omega ^{2}L_{X})^{-1}e^{ik}-C_{X}e^{-ik},  \label{ffgg}
\end{align}%
describing the imaginary superconducting pairing model $H^{y}$.

(ii) 
In the case of the circuit in Fig.\ref{FigCircuit}(b) we explicitly obtain (\ref{MatrixJ}), where
$f_{1}$ and $f_{2}$ are given by (\ref{ffgg}) and 
\begin{equation}
g_{1}=g_{2}=(i\omega R_{b})^{-1}e^{ik}-(i\omega R_{f})^{-1}e^{-ik},
\end{equation}%
describing the real superconducting pairing model $H^{x}$.

The key procedure is to equate the circuit Laplacian (\ref{MatrixJ}) with
the Hamiltonian\ (\ref{KitaHami}). In so doing, it is necessary to require
the PHS for the circuit, which requires us to tune the parameters to satisfy 
$\omega _{0}\equiv 1/\sqrt{LC}=1/\sqrt{L_{0}C_{0}}=1/\sqrt{L_{X}C_{X}}$, and
set the AC frequency as $\omega =\omega _{0}$. At this frequency, we may set 
$J_{ab}\left( \omega \right) =i\omega H_{ab}^{i}\left( \omega \right) $,
which dictates the correspondence between the circuit and the superconductor 
model as $t=-C$, $\mu =-2C+C_{0}$, $\Delta _{y}=C_{X}$, and $\Delta
_{x}=(\omega _{0}R_{X})^{-1}$.

The system is precisely at the topological phase-transition point $%
\left\vert \mu \right\vert =\left\vert 2t\right\vert $ without the
capacitors $C_{0}$ and the inductors $L_{0}$, since the condition $\mu =-2t$
is satisfied. It is topological in the presence of $C_{0}$ and $L_{0}$. The
system turns into a trivial phase when we exchange the capacitors $C_{0}$
and inductors $L_{0}$ connected to ground.

By calculating the LDOS as in Fig.\ref{FigKitaLDOS}(a1), 
we find the emergence of the zero-energy edge states in the topological phase.
They are observable by measuring the impedance between the $a$ and $b$ nodes, which is
given by\cite{Hel} $Z_{ab}\equiv V_{a}/I_{b}=G_{ab}$, where $G$ is the Green function
defined by the inverse of the Laplacian $J$, $G\equiv J^{-1}$. We show
numerical results in Fig.\ref{FigKitaLDOS}(a2) for typical values of
parameters, where we have set one node at the left or right edge. The
behavior of the impedance is very similar to that of the LDOS.

(iii) We next consider the $p_{x}+ip_{y}$ model, whose Hamiltonian is given
by (\ref{KitaHami}) with 
\begin{align}
H_{t}& =t(\cos k_{x}+\cos k_{y})-\mu ,  \label{Ht} \\
H_{\text{SC}}& =\Delta (\sigma _{x}\sin k_{x}+\sigma _{y}\sin k_{y}).
\label{HSO}
\end{align}%
The model is simulated by layering the circuit for $H^{y}$ in the $x$
direction and the circuit for $H^{x}$ in the $y$ direction as shown in Fig.\ref{FigCircuit}(d). 
The relations between the parameters are given by
\begin{equation}
t=-C,\quad \mu =-2C+C_{0},\quad \Delta =C_{X}=(\omega _{0}R_{X})^{-1}.
\label{SysParam}
\end{equation}%
It is a topological superconductor for $\left\vert \mu \right\vert
<\left\vert 2t\right\vert $, where the zero-energy edge states emerge along all
four edges.

\textit{Braiding of Majorana corner states: }
The zero-energy edge states are Majorana states in the Kitaev model, 
but it is not easy to study the braiding\cite{AliceaBraid} in the electric-circuit formalism. 
The zero-energy edge states in $p_{x}+ip_{y}$ model is 2D. 
We proceed to investigate a 2D model possessing a pair of zero-energy corner states to explore the braiding.

\begin{figure}[t]
\centerline{\includegraphics[width=0.48\textwidth]{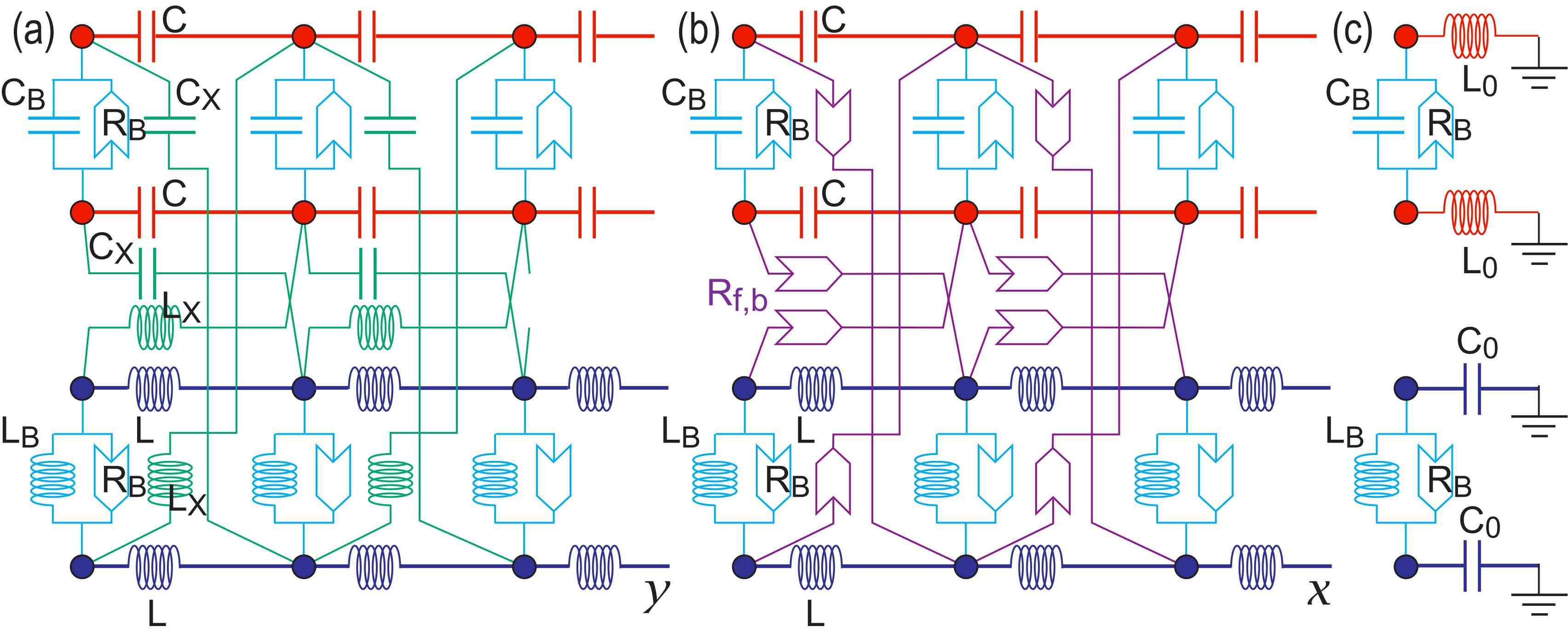}}
\caption{Electric-circuit realization of the BHZ model together with
effective field $B$. The $B$ field is simulated by parts involving ($C_B,
R_B $) and ($L_B,R_B$) colored in cyan. (a) 2D view into the $x$ direction.
(b) 2D view into the $y$ direction. (c) Each node is connected to the ground
by an inductor or a capacitor in (a) and (b). We have set $t=1,\protect\mu %
=1,\Delta =1$ and $B=1/2$.}
\label{FigBHZCircuit}
\end{figure}

Such a model is given by the Bernevig-Hughes-Zhang Hamiltonian\cite{BHZ}, $%
H^{\text{BHZ}}=H_{t}\tau _{z}+H_{\text{SO}}\tau _{x}$, where $H_{t}$ and $H_{%
\text{SO}}$ are given by (\ref{Ht}) and (\ref{HSO}), respectively. Although
it is proposed for a topological insulator, it has the PHS, $\Xi
^{-1}H\left( \mathbf{k}\right) \Xi =-H\left( -\mathbf{k}\right) $ with $\Xi
=\tau _{y}\sigma _{y}K$, where $K$ represents complex conjugation. When the
Zeeman term $H_{Z}=B\left( \sigma _{x}\cos \theta +B\sigma _{y}\sin \theta
\right) $ is applied, it becomes a second-order topological superconductor
with the emergence of zero-energy topological corner states\cite%
{Zhu,TopoSwitch}. However, it breaks the PHS\cite{YWang}. Here we propose
the term $H_{\tau Z}=B\tau _{z}\left( \sigma _{x}\cos \theta +\sigma
_{y}\sin \theta \right) $, which respects the PHS. Such a term does not
exist in condensed matter, but it is allowed in electric circuits: See cyan
parts in Fig.\ref{FigBHZCircuit}.

We simulate the BHZ model with the $B$ field, $H_{\tau Z}^{\text{BHZ}%
}(\theta )\equiv H^{\text{BHZ}}+H_{\tau Z}$, by an electric circuit.
It is straightforward to generalize the above circuits in Fig.\ref{FigCircuit} to the present one as in Fig.\ref{FigBHZCircuit}. 
Since it is a four-band model, we use four main wires.
By analyzing the Kirchhoff current law, we may derive the circuit
Laplacian $J_{ab}\left( \omega \right) $, which is now a $4\times 4$ matrix.
Solving $J_{ab}\left( \omega \right) =i\omega H_{\tau Z}^{\text{BHZ}}$, we
obtain the correspondence between the system parameters.
They are given by (\ref{SysParam}) supplemented by $B_{x}=C_{B}$ or $L_{B}$ and $B_{y}=R_{B}$ for the $B$ field.
We show the LDOS and the impedance at $\theta =\pi /4$, $5\pi /4$ in Fig.\ref{FigBHZdos}(a)--(c), 
where the zero-energy corner states are clearly observed. 
See the LDOS at other values of $\theta $ elsewhere.

\begin{figure}[t]
\centerline{\includegraphics[width=0.48\textwidth]{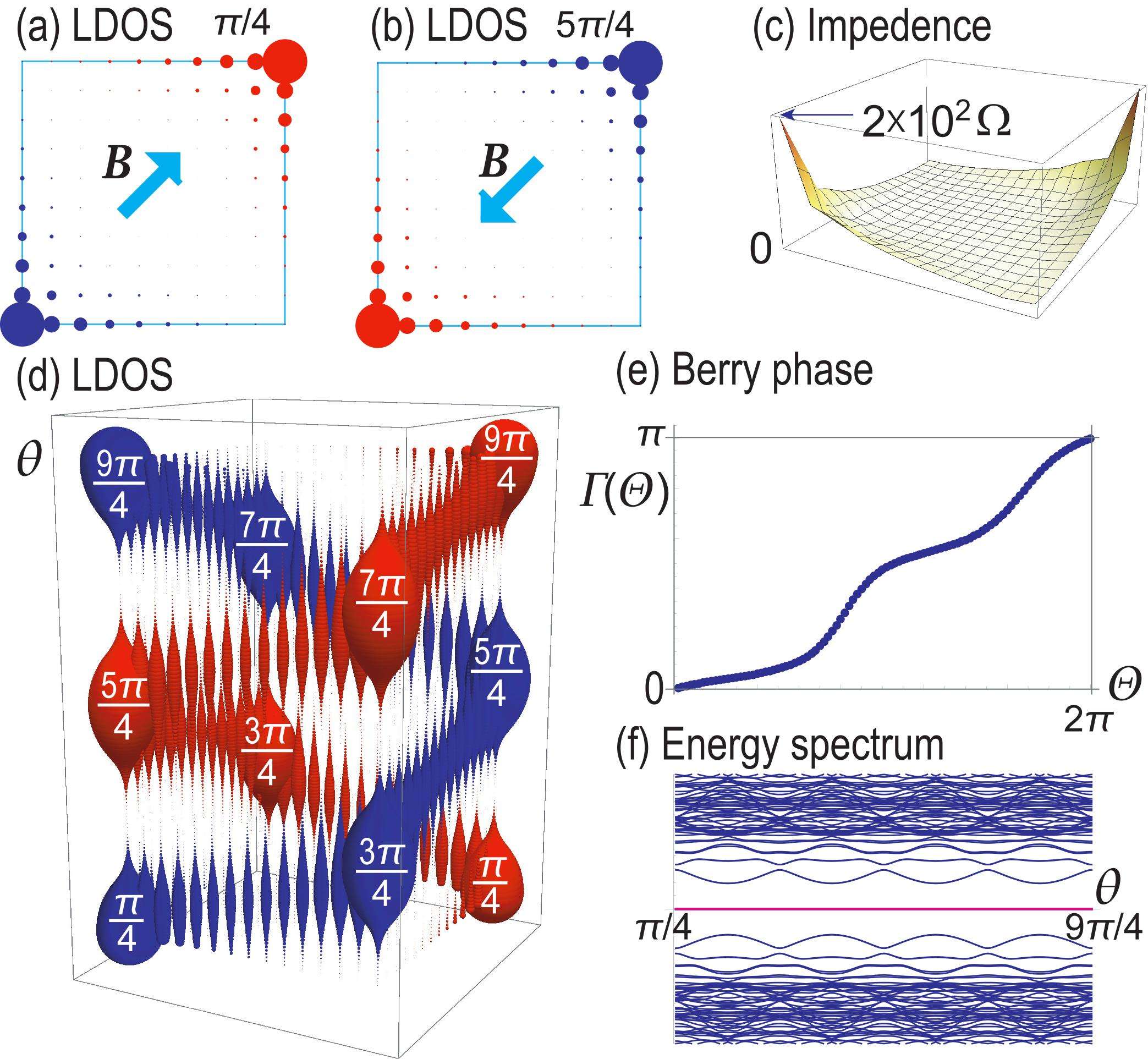}}
\caption{LDOS of the corner states (a) at $\protect\theta =\protect\pi /4$
and (b) at $\protect\theta =5\protect\pi /4$. These two are related by a
single-exchange operation. Cyan arrows show the direction of the $B$ field.
(c) Impendence is the same at $\protect\theta =\protect\pi /4$ and $5\protect%
\pi /4$. (d) LDOS of the corner states for $\protect\pi /4<\protect\theta <9%
\protect\pi /4$, and their braiding. (e) Evolution of the Berry phase
towards a double-exchange operation. (f) Energy spectrum evolution during
the braiding. The Majorana corner states in red remains to be
separated from the bulk spectrum in blue. }
\label{FigBHZdos}
\end{figure}

The zero-energy corner states subject to the PHS are Majorana states in condensed matter physics.
A key question is whether they may be called Majorana states in electric circuits.
Majorana particles are known to be Ising anyons possessing the property $%
\sigma ^{2}=-1$, where $\sigma $ denotes the single-exchange operation\cite%
{TQC,Kitaev}. It is well known that only fermions and bosons are possible in
3D, for which $\sigma ^{2}=1$. On the other hand, anyons are possible only in 2D.
We recognize that this anyonic property is most important 
as a characteristics of Majorana states for future application to quantum computers.

We investigate the braiding for a pair of corner states. By increasing $%
\theta $ continuously from $\pi /4$ to $5\pi /4$ ($9\pi /4$), we can
exchange the position of two corner states once (twice) as in Fig.\ref%
{FigBHZdos}(d). The corner states remain to be zero-energy states with a
finite gap during the process, as shown in Fig.\ref{FigBHZdos}(f). Thus,
they remain well separated from the bulk bands during the exchange of the
two corner states. The key property is how the wave function changes as $%
\theta $ increases. Note that the phase of the wave function is observable
by the phase shift in the electric circuit.

Let $\left\vert \psi _{\alpha }(\theta )\right\rangle $ be an eigenstate of $%
H_{\tau Z}^{\text{BHZ}}(\theta )$. When we increase $\theta $ adiabatically,
the wave function $\left\vert \psi _{\alpha }(\theta )\right\rangle $
develops as\cite{TQC,Kitaev,DasSarma}%
\begin{equation}
\left\vert \psi _{\alpha }\left( \theta \right) \right\rangle =\sum_{\beta
=1,2}e^{i\Gamma _{\alpha \beta }(\Theta )}\left\vert \psi _{\beta }\left(
\pi /4\right) \right\rangle ,
\end{equation}%
where $\Theta =\theta -\pi /4$, and $\Gamma _{\alpha \beta }\left( \Theta
\right) $ is the Berry phase,%
\begin{equation}
\Gamma _{\alpha \beta }\left( \Theta \right) =i\int_{\pi /4}^{\pi /4+\Theta
}\left\langle \psi _{\alpha }(\theta )\right\vert \partial _{\theta
}\left\vert \psi _{\beta }(\theta )\right\rangle d\theta .
\end{equation}%
There are two-fold degenerate zero-energy corner states at $\theta =\pi /4$.
Since these two states are well separated as in Fig.\ref{FigBHZdos}(a), we
may label them by $\alpha =1,2$. Furthermore, we may construct the
eigenfunctions continuous in $\theta $ such that $\left\langle \psi _{\alpha
}(\theta )\right\vert \psi _{\beta }(\theta )\rangle =\delta _{\alpha \beta
} $ for any value of $\theta $. Then, it follows that $\Gamma _{\alpha \beta
}\left( \Theta \right) $ is diagonal; $\Gamma \left( \Theta \right) \equiv
\Gamma _{11}\left( \Theta \right) =\Gamma _{22}\left( \Theta \right) $. We
show a numerical result for $\Gamma \left( \Theta \right) $ in Fig.\ref%
{FigBHZdos}(e). In particular, we obtain $\Gamma (2\pi )=\pi $.

The single and double exchanges correspond to the rotations $\Theta =\pi $
and $\Theta =2\pi $, respectively. After the double exchange we obtain $%
\Gamma _{\alpha \beta }(2\pi )=\pi \delta _{\alpha \beta }$, which yields $%
\left\vert \psi _{1}\right\rangle \rightarrow -\left\vert \psi
_{1}\right\rangle $ and $\left\vert \psi _{2}\right\rangle \rightarrow
-\left\vert \psi _{2}\right\rangle $, or $\sigma ^{2}=-1$. Consequently, the
braiding is satisfied by the corner states in electric circuits. 
Hence, let us call the zero-energy corner states Majorana states also in the electric-circuit formalism.
We can check that the braiding is robust against disorders.

\textit{Non-Hermitian Majorana states}: We have so far analyzed Hermitian
models. It is easy to make a model non-Hermitian by introducing resistors
and diodes into an electric circuit\cite{EzawaLCR,EzawaSkin}. Since we are
interested in Majorana states, we only consider the systems respecting the
PHS. The PHS is generalized straightforwardly to the non-Hermitian case.
The Majorana particle is identical to its antiparticle. A new
feature is that its energy is zero or pure imaginary.

A non-Hermitian model is either reciprocal or nonreciprocal. To make the
analysis concrete, we explicitly consider a non-Hermitian Kitaev model
respecting the PHS. The Hamiltonian (\ref{KitaHami}) together
with (\ref{KitaHop}) and (\ref{KitaSO}) is generalized as
\begin{equation}
H\left( k\right) =i\gamma \mathbb{I}+\left( 
\begin{array}{cc}
f(t^{b},t^{f};k) & g(\Delta ^{b},\Delta ^{f};k) \\ 
g(\Delta ^{b\ast },\Delta ^{f\ast };k) & -f(t^{b\ast },t^{f\ast };k)%
\end{array}%
\right) , \label{KitaNonH}
\end{equation}%
with $f(t^{b},t^{f};k)=t^{b}e^{ik}+t^{f}e^{-ik}-\mu $, $g(\Delta ^{b},\Delta
^{f};k)=-i\left( \Delta ^{b}e^{ik}-\Delta ^{f}e^{-ik}\right) $, and $\gamma $
representing dissipation, where $t^{b}$ ($t^{f}$) is a backward (forward)
hopping amplitude, $\Delta ^{b}$ ($\Delta ^{f}$) is a backward (forward)
superconducting pairing amplitude, and $\mu $ is the chemical potential.
Parameters $t^{b}$, $t^{f}$, $\Delta ^{b}$ and $\Delta ^{f}$ take complex
values, while $\mu $ and $\gamma $ take real values. It satisfies the PHS, $%
\Xi ^{-1}H\left( k\right) \Xi =-H\left( -k\right) $, with $\Xi =\sigma _{x}K$%
.

By diagonalizing (\ref{KitaNonH}), the bulk gap is found to close at
\begin{equation}
\left\vert \mu \right\vert =\left\vert \text{Re}\left( t^{b}+t^{f}\right)
\pm |\Delta ^{b}-\Delta ^{f}|\right\vert .  \label{GapB}
\end{equation}%
Gap-closing points are not phase-transition points 
when skin edge states are present in non-Hermitian theory\cite%
{Xiong,Yao,Yao2,Kunst,Lee,Jin,TopSkin,EzawaLCR}.

We seek for topological phases. The system remains to be in the class Z$_{2}$%
\ even for the non-Hermitian system\cite{KawabataST}, and hence, the
topological number is given by the Z$_{2}$ invariant $\nu $ in the original
Kitaev model as 
\begin{equation}
\left( -1\right) ^{\nu }=-\text{sgn}\left[ H_{z}\left( 0\right) H_{z}\left(
\pi \right) \right] ,  \label{Z2B0}
\end{equation}%
where 
$H_{z}$ is the coefficient of $\sigma _{z}$ by expanding $H$ as $H\left( k\right)
=\sum_{\alpha =0,x,y,z}H_{\alpha }\left( k\right) \sigma _{\alpha }$, and 
$k=0$, $\pi $ are the PHS invariant momenta.
The formula (\ref{Z2B0}) is valid also for the non-Hermitian system because of
the PHS. Calculating it explicitly we find that
\begin{equation}
\left( -1\right) ^{\nu }=\text{sgn}\left[ \left[ \text{Re}\left(
t^{b}+t^{f}\right) \right] ^{2}-\mu ^{2}\right] .  \label{Z2B}
\end{equation}%
It follows from (\ref{Z2B}) that there are two phases with the phase-transition points 
$\mu _{\pm }=\pm \left\vert \text{Re}\left( t^{b}+t^{f}\right) \right\vert $.
The system is topological ($\nu =1$) for $\left\vert \mu \right\vert
<\left\vert \text{Re}\left( t^{b}+t^{f}\right) \right\vert $ and trivial ($%
\nu =0$) for $\left\vert \mu \right\vert >\left\vert \text{Re}\left(
t^{b}+t^{f}\right) \right\vert $. 
The justification of (\ref{Z2B}) as the topological number is made 
by confirming numerically the non-Hermitian bulk-edge correspondence\cite{Yao,Yao2,Kunst,Lee,Jin,EzawaLCR}. 
The gap-closing points (\ref{GapB}) become identical to the topological phase-transition points $\mu _{\pm }$ 
when the system is reciprocal,  $\Delta ^{b}=\Delta ^{f}$.

The LDOS is shown for all eigen-energies in Fig.\ref{FigKitaLDOS}(a1)--(c1)
by taking typical values of sample parameters in the Hermitian, reciprocal
non-Hermitian and nonreciprocal non-Hermitian cases. The characteristic
feature of the nonreciprocal non-Hermitian model is the emergence of skin
edge states as in [Fig.\ref{FigKitaLDOS}(c1)], where all the eigen states
are localized at one edge, as was first found in the non-Hermitian
Su-Schrieffer-Heeger model\cite{Yao,Yao2,Kunst}. Namely, two topological edge
states are mixed between themselves and furthermore they are mixed with the
bulk states in the vicinity of one edge. 
This is also the case for the topological corner states in the nonreciprocal non-Hermitian BHZ model.
The braiding becomes meaningless in the nonreciprocal non-Hermitian models
because there are no separated corner states.

We may construct a reciprocal non-Hermitian model respecting
the PHS, by inserting a resistor $R$ to a capacitance $C$ and an inductor $L$
in series in Fig.\ref{FigBHZCircuit}. The circuit Laplacian is obtained just
by replacing $C\rightarrow (1/C+i\omega R)^{-1}$ and $\omega
^{2}L\rightarrow \omega ^{2}L-i\omega R$. We can check that the
resultant non-Hermitian model is different from the original Hermitian model
only by a pure imaginary shift. In such a case the braiding holds just as it
is since the wave functions are not modified.

\textit{Discussion:} We have shown that Majorana edge and corner states are
simulated by electric circuits. The Majorana corner states emerge in
electric circuits whose size is as small as $N=4$, which is a
benefit on future high-density applications. Majorana states will be a key
for future topological quantum computations, where the essential property is
the braiding between two Majorana states. Furthermore, various extensions
are possible to electric circuits for such as the dimerized Kitaev model\cite%
{DimKitaev} and the Kitaev ladder model\cite{Ladder}. Our results might open
a new way for topological quantum computations based on Majorana states in
electric circuits.

The author is very grateful to A. Kurobe for fruitful conversations on the
subject, which motivated the present work. He is also thankful to Y. Tanaka
and N. Nagaosa for helpful discussions on the subject. He also thanks to T.
Pahomi and A. A. Soluyanov on discussions on the braiding of Majorana
states. This work is supported by the Grants-in-Aid for Scientific Research
from MEXT KAKENHI (Grants No. JP17K05490, No. JP15H05854 and No.
JP18H03676). This work is also supported by CREST, JST (JPMJCR16F1 and
JPMJCR1874).

\end{document}